\documentstyle[aps,prd,preprint,floats,epsfig,tighten]{revtex}

\newcommand{\NP}[1]{ Nucl.\ Phys.\ {\bf #1}}
\newcommand{\ZP}[1]{ Z.\ Phys.\ {\bf #1}}

\newcommand{\PL}[1]{ Phys.\ Lett.\ {\bf #1}}

\newcommand{\PRep}[1]{Phys.\ Rep.\ {\bf #1}}
\newcommand{\PR}[1]{Phys.\ Rev.\ {\bf #1}}
\newcommand{\PRL}[1]{ Phys.\ Rev.\ Lett.\ {\bf #1}}

\newcommand{\Pom}{${\cal P}$}
\newcommand{\Odd}{${\cal O}$}
\newcommand{\pom}{{\cal P}}
\newcommand{\odd}{{\cal O}}
\newcommand{\etal} {{\em et al.}}

\begin{document}

\begin{flushright}
SLAC-PUB-8095\\
April 1999\\
Rev. June 1999
\end{flushright}
\vfill
\begin{center}
{\LARGE {Odderon-Pomeron Interference}\footnote{\baselineskip=13pt
Research partially supported by the Department of Energy under contract
DE--AC03--76SF00515, the Spanish CICYT under contract AEN96--1673,
and the Swedish Natural Science Research Council,
contract F--PD 11264--301.}}

\vspace{15mm}
{\bf Stanley J. Brodsky and Johan Rathsman}\\
\vspace{5mm}
{\em Stanford Linear Accelerator Center \\
Stanford University, Stanford, California 94309}

\vspace{15mm}
{\bf Carlos Merino}\\
\vspace{5mm}
{\em Departamento de F\'{\i}sica de Part\'{\i}culas \\
 Universidade de Santiago de Compostela \\
15706 Santiago de Compostela, Spain}
\end{center}

\begin{abstract}
We show that the asymmetry in the fractional energy of charm versus
anticharm jets  produced in high energy diffractive photoproduction is
sensitive to the interference of the Odderon $(C = -)$  and Pomeron
$(C = +)$ exchange amplitudes in QCD. We predict the dynamical shape
of the asymmetry in a simple model and estimate its magnitude to be
of the order 15\% using an Odderon coupling to the proton which
saturates constraints from proton-proton vs.~proton-antiproton elastic 
scattering. Measurements of this asymmetry at HERA could provide firm 
experimental evidence for the presence of Odderon
exchange in the high energy limit of strong interactions.
\end{abstract}
\vfill
\centerline{Submitted to Physics Letters B.}
\vfill
\newpage

The existence of odd charge-conjugation,  zero flavor-number exchange
contributions to high energy hadron scattering
amplitudes is a basic prediction of quantum chromodynamics, following
simply from the existence of the color-singlet exchange
of three reggeized gluons in the
$t-$channel~\cite{kwiencinski_bartels}.
In Regge theory, the ``Odderon" contribution  is dual to a sum over
$C = P= -1$ gluonium states in the $t$-channel~\cite{lukaszuk_nicolescu}.
In the case of reactions
which involve high momentum transfer, the deviation of the Regge
intercept of the Odderon trajectory from $\alpha_{\cal O}(t=0) = 1$ can
in principle be computed~\cite{Lipatov2,Braun,Wosieck,Gauron2}
from perturbative QCD in
analogy to the methods used to compute the properties of the hard BFKL
Pomeron~\cite{BFKL}. (For a more complete history of the Odderon we refer
the reader to~\cite{landshoff_nachtmann} and~\cite{nicolescu} and
references therein.)

In the case of low momentum transfer reactions, the Odderon exchange
amplitude should yield a roughly energy-independent contribution to the
difference of proton-proton vs.~proton-antiproton cross sections.  It
should also be seen in high energy diffractive pseudoscalar meson
photoproduction, such as
$\gamma p \to \pi^0 p$~\cite{pseudo,rueter,kilian_nachtmann} and
$\gamma \gamma \to \pi^0 \pi^0$~\cite{ginzburg},
since these amplitudes demand odd $C$ exchange.
Despite these theoretical expectations, there is as yet no firm
experimental evidence for any Odderon contribution in the high-energy
limit $s\gg|t|$. A hint of the Odderon was seen in ISR results~\cite{ISR}
($\sqrt{s}=52.8$ GeV) in the difference between the elastic $pp$ and 
$p\bar{p}$ differential cross-sections at the diffractive minimum, 
$t\sim-1.3$ GeV$^2$. A realization of the Odderon in perturbation 
theory is represented by the Landshoff contribution to large angle $pp$
scattering~\cite{landshoff}.

Recent results from the electron-proton collider experiments at
HERA~\cite{HERA}---the rapidly-rising behavior of proton
structure functions at small $x$, the rapidly-rising
diffractive vector meson electroproduction rates, and the steep rise of
the
$J/\psi$ photoproduction cross section have brought  renewed interest in
the nature and behavior of the  Pomeron
in QCD (see for example~\cite{Brodsky1,CKMT1}). In this letter we
propose an experimental test well suited to HERA kinematics which should
be able to disentangle the contributions of both the Pomeron and the
Odderon to  diffractive production of charmed jets. By forming a charge
asymmetry in the energy of the charmed jets, we can determine the
relative importance of the Pomeron ($C=+$) and the Odderon ($C=-$)
contributions, and their interference, thus providing a new
experimental test of the separate existence of these two objects.
Since the asymmetry measures the
Odderon amplitude linearly, even a relatively weakly-coupled amplitude
should be detectable.

\begin{figure}[htb]
\begin{center}
\leavevmode
\epsfbox{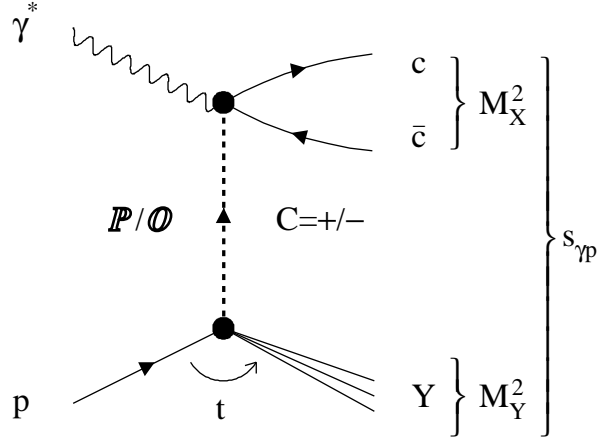}
\end{center}
\caption[*]{The amplitude for the diffractive process
$\gamma p \to c\bar{c} Y$ with Pomeron (\Pom) or Odderon (\Odd)
exchange.}
\label{fig:amplitude}
\end{figure}

Consider the diagrams in Fig.~\ref{fig:amplitude}
describing the amplitude for diffractive
photoproduction of a charm quark anti-quark pair.
The leading diagram is given by single Pomeron exchange~(two reggeized
gluons), and the  next term in the Born expansion is given by the exchange
of one  Odderon~(three reggeized gluons). In the following we will
focus on the situation when the diffractively scattered
proton $p^\prime$ stays intact; however, the formulae will be equally
valid when  the diffractively scattered proton
is excited to a low mass system $Y$. We only require the
invariant  mass of the system $M_Y^2$ to be small compared to the
invariant mass of the
$c\bar{c}$ pair $M_X^2$. In fact, as pointed out by Rueter
\etal~\cite{rueter}, the cross-section for the diffractively excited protons
can be significantly larger than the elastic cross-section 
in specific models such as diquark clustering in the proton.
The virtuality of the incoming
photon $Q^2$ can be zero or small since the invariant mass of the
$c\bar{c}$ pair $M_X^2$ is large. Thus we are considering both
diffractive photoproduction and leptoproduction, although in the
following we will specialize to the case of photoproduction for
which the rate observed at HERA is much larger. Our results can easily be
generalized to non-zero $Q^2$.

The total
center of mass energy of the $\gamma p$ system will be denoted $s_{\gamma
p}$  which should be distinguished from the total $ep$ cms energy.
Denoting the photon momentum by $q$, the proton momentum by $p$, and
the momenta of the charm quark (antiquark) by $p_c$ ($p_{\bar{c}}$),
the energy sharing of the $c\bar{c}$ pair is given by the variable
\begin{equation}
z_{c(\bar{c})}=\frac{p_{c(\bar{c})}p}{qp}=\frac{E_{c(\bar{c})}}{E_{\gamma^*}}
\end{equation}
where the latter equality is true in the proton rest frame.
It follows that $z_{c}+z_{\bar{c}}=1$
in Born approximation at the parton level. The finite charm quark mass
restricts the range of $z$ to
\begin{equation}
\frac{1}{2}-\sqrt{\frac{1}{4}-\frac{m_c^2}{M_X^2}} \leq z
\leq \frac{1}{2}+\sqrt{\frac{1}{4}-\frac{m_c^2}{M_X^2}}
\end{equation}
where $M_X^2$ is the invariant mass of the $c\bar{c}$ pair which
is related to the total $\gamma p$ cms energy $s_{\gamma p}$ by
\begin{equation}
M_X^2=({\xi}p+q)^2\simeq2{\xi}pq \simeq{\xi}s_{\gamma p}
\end{equation}
where $\xi$ is effectively the longitudinal momentum
fraction of the proton carried by the Pomeron/Odderon and the proton mass
is neglected.

Regge theory, which is applicable in the kinematic region
$s_{\gamma p} \gg M_X^2 \gg M_Y^2$,
together with crossing symmetry, predicts the phases and analytic
form of high energy amplitudes (see, for example,
Refs.~\cite{collins} and~\cite{kaidalov}).
The amplitude for the diffractive process
$\gamma p \to c\bar{c} p^\prime$ with Pomeron (\Pom) or Odderon (\Odd)
exchange can be written as
\begin{eqnarray}\label{eq:ampl}
{\cal M}^{\pom/\odd}(t,s_{\gamma p},M_X^2,z_c) & \propto &
 g_{pp^\prime}^{\pom/\odd}(t)
\left(\frac{s_{\gamma p}}{M_X^2}\right)^{\alpha_{\pom/\odd}(t)-1}
\nonumber \\
&& \times
\frac{\left(1+S_{\pom/\odd}e^{-i\pi\alpha_{\pom/\odd}(t)}\right)}
{\sin\pi\alpha_{\pom/\odd}(t)}
g_{\pom/\odd}^{\gamma c\bar{c}}(t,M_X^2,z_c)
\end{eqnarray}
where $S_{\pom/\odd}$ is the signature\footnote{Even (odd) signature
corresponds to an exchange which is (anti)symmetric under the interchange
$s\leftrightarrow u$.} which is $+(-)1$ for the Pomeron (Odderon).
In the Regge approach the upper vertex
$g_{\pom/\odd}^{\gamma c\bar{c}}(t,M_X^2,z_c)$
can be treated as a
local real coupling such that the phase is contained in the signature
factor.  In the same way the factor $g_{pp^\prime}^{\pom/\odd}(t)$
represents the lower vertex. For our purposes it will be convenient
to rewrite the signature factor in the following way,
\begin{equation}
\frac{\left(1+S_{\pom/\odd}e^{-i\pi\alpha_{\pom/\odd}(t)}\right)}
{\sin\pi\alpha_{\pom/\odd}(t)}
= \left\{ \begin{array}{ll}
{\displaystyle \frac{\cos{ \frac{\pi\alpha_\pom(t)}{2} } -
i \sin { \frac{\pi\alpha_\pom(t)}{2}}}
{\sin { \frac{\pi\alpha_\pom(t)}{2}}}}
 &  \quad \mbox{for $S_{\pom}=1$} \\ \\
{\displaystyle \frac{\sin{ \frac{\pi\alpha_\odd(t)}{2}}+
i\cos{ \frac{\pi\alpha_\odd(t)}{2}}}
{\cos{ \frac{\pi\alpha_\odd(t)}{2}}}} &  \quad \mbox{for $S_{\odd}=-1$}
\end{array} \right. .
\label{eq:polar}
\end{equation}
In the literature it has become customary to absorb
the pole factors $1/\sin{\frac{\pi\alpha_\pom(t)}{2}}$ and
$1/\cos{\frac{\pi\alpha_\odd(t)}{2}}$ into the couplings
$\left(g_{pp^\prime}^{\pom/\odd}(t)\right)^2$, but we will keep them explicit
since we want to treat the upper and lower vertex separately.

\begin{figure}[htb]
\begin{center}
\leavevmode
\epsfbox{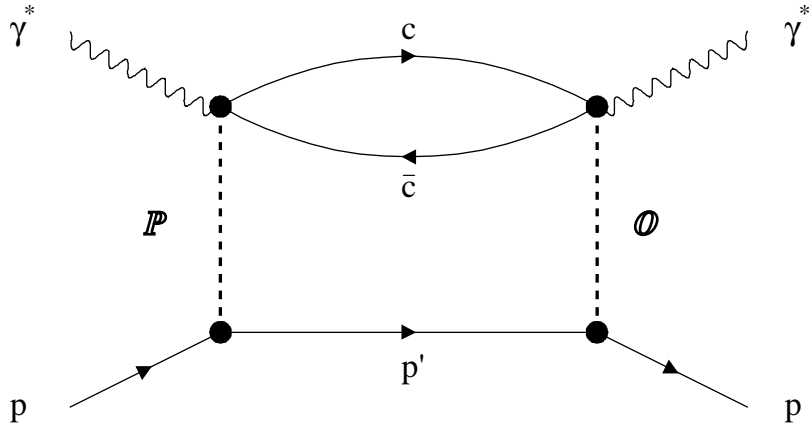}
\end{center}
\caption[*]{The interference between Pomeron (\Pom) or Odderon (\Odd)
exchange in the diffractive process $\gamma p \to c\bar{c} p^\prime$.}
\label{fig:interference}
\end{figure}

In general the Pomeron and Odderon exchange
amplitudes will  interfere, as illustrated in Fig.~\ref{fig:interference}.
The contribution of the
interference term to the total  cross-section is  zero, but it
does contribute to charge-asymmetric rates. Thus we propose to study
photoproduction of $c$-$\bar{c}$ pairs and measure the asymmetry in
the energy fractions $z_c$ and $z_{\bar{c}}$.  More generally,
one can use other charge-asymmetric kinematic configurations, as well as
bottom or strange quarks.

Given the amplitude ~(\ref{eq:ampl}),
the contribution to the cross-section from the
interference term depicted in Fig.~\ref{fig:interference}
is proportional to
\begin{eqnarray} \label{eq:int}
\frac{d\sigma^{int}}{dtdM_X^2dz_c} & \propto &
{\cal M}^{\pom}(t,s_{\gamma p},M_X^2,z_c)
\left\{{\cal M}^{\odd}(t,s_{\gamma p},M_X^2,z_c) \right\}^\dagger +h.c.
\nonumber \\[1ex] & = &
 g_{pp^\prime}^{\pom}(t)g_{pp^\prime}^{\odd}(t)
\left(\frac{s_{\gamma p}}{M_X^2}\right)^{\alpha_{\pom}(t)+\alpha_{\odd}(t)-2}
\frac{2\sin \left[ \frac{\pi}{2}\left(\alpha_{\odd}(t)-\alpha_{\pom}(t)\right)
 \right]}
{\sin{ \frac{\pi\alpha_\pom(t)}{2}} \cos{ \frac{\pi\alpha_\odd(t)}{2}}}
\nonumber \\[1ex] && \times
g_{\pom}^{\gamma c\bar{c}}(t,M_X^2,z_c)
g_{\odd}^{\gamma c\bar{c}}(t,M_X^2,z_c) \ .
\end{eqnarray}
In the same way we obtain the contributions to the cross-section from the
non-interfering terms for Pomeron and Odderon exchange,
\begin{eqnarray} \label{eq:sqr}
\frac{d\sigma^{\pom/\odd}}{dtdM_X^2dz_c} & \propto &
 \left\{ \begin{array}{ll}
{\displaystyle \left[ g_{pp^\prime}^{\pom}(t)
\left(\frac{s_{\gamma p}}{M_X^2}\right)^{\alpha_{\pom}(t)-1}
g_{\pom}^{\gamma c\bar{c}}(t,M_X^2,z_c)
/\sin\frac{\pi\alpha_\pom(t)}{2}
\right]^2 }
 &  \quad \mbox{for $S_{\pom}=1$} \\ \\
{\displaystyle \left[ g_{pp^\prime}^{\odd}(t)
\left(\frac{s_{\gamma p}}{M_X^2}\right)^{\alpha_{\odd}(t)-1}
g_{\odd}^{\gamma c\bar{c}}(t,M_X^2,z_c)
/\cos\frac{\pi\alpha_\odd(t)}{2}\right]^2 }
 &  \quad \mbox{for $S_{\odd}=-1$}
\end{array} \right.  \; .
\end{eqnarray}
We note the different charge conjugation properties of the
upper vertices:
\begin{eqnarray} \label{eq:cc}
 g_{\pom}^{\gamma c\bar{c}}(t,M_X^2,z_c) & = &
 - g_{\pom}^{\gamma c\bar{c}}(t,M_X^2,z_{\bar{c}})
\nonumber \\
 g_{\odd}^{\gamma c\bar{c}}(t,M_X^2,z_c) & = &
                       g_{\odd}^{\gamma c\bar{c}}(t,M_X^2,z_{\bar{c}})
     	\; .
\end{eqnarray}
The interference term can then be isolated by forming
the charge asymmetry,
\begin{equation} \label{eq:asym}
{\cal A}(t,M_X^2,z_c) =
\frac{\displaystyle \frac{d\sigma}{dtdM_X^2dz_c}
                  - \frac{d\sigma}{dtdM_X^2dz_{\bar{c}}} }
{\displaystyle \frac{d\sigma}{dtdM_X^2dz_c}
             + \frac{d\sigma}{dtdM_X^2dz_{\bar{c}}} } \; .
\end{equation}
Inserting Eqs.~(\ref{eq:int}), (\ref{eq:sqr}) and (\ref{eq:cc}) into
Eq.~(\ref{eq:asym}) then gives the predicted asymmetry,
\begin{eqnarray}\label{eq:pred}
{\cal A}(t,M_X^2,z_c) & = & \frac{
{\displaystyle g_{pp^\prime}^{\pom}
g_{pp^\prime}^{\odd}
\left(\frac{s_{\gamma p}}{M_X^2}\right)^{\alpha_{\pom}+\alpha_{\odd} }
\frac{2\sin \left[\frac{\pi}{2}\left(\alpha_{\odd}-\alpha_{\pom}\right)
 \right]}
{\sin{ \frac{\pi\alpha_\pom}{2}} \cos{ \frac{\pi\alpha_\odd}{2}}}
g_{\pom}^{\gamma c\bar{c}}
g_{\odd}^{\gamma c\bar{c}}}
}
{{\displaystyle \left[g_{pp^\prime}^{\pom}
\left(\frac{s_{\gamma p}}{M_X^2}\right)^{\alpha_{\pom} }
g_{\pom}^{\gamma c\bar{c}}
/\sin\frac{\pi\alpha_\pom}{2}
\right]^2  +
\left[g_{pp^\prime}^{\odd}
\left(\frac{s_{\gamma p}}{M_X^2}\right)^{\alpha_{\odd} }
g_{\odd}^{\gamma c\bar{c}}
/\cos\frac{\pi\alpha_\odd}{2}
\right]^2 }}
\end{eqnarray}
where  the arguments have been dropped for clarity.
This is the general form of the Pomeron-Odderon interference
contribution in Regge theory. In the following we will
give numerical estimates for the different components and also
calculate the asymmetry using the Donnachie-Landshoff
model for the Pomeron\cite{donnachie_landshoff}.

The functional dependence of the asymmetry on the kinematical variables can be
obtained by varying the kinematic variables one at a time. In this way it will
be possible to obtain new information about Odderon exchange in relation to
Pomeron exchange. Furthermore, we expect the main dependence in the different
kinematic variables to come from different factors in the asymmetry. For
instance, the invariant mass $M_X$ dependence is mainly given by the power
behavior, $\left({s_{\gamma p}}/{M_X^2}\right)^{ \alpha_{\odd}(t)
-\alpha_{\pom}(t)}$, and it will thus  provide direct information about the
difference between $\alpha_{\odd}$ and $\alpha_{\pom}$. Another interesting
question which can be addressed from observations of the asymmetry is  the
difference in the $t$-dependence of $g_{pp^\prime}^{\odd}$ and
$g_{pp^\prime}^{\pom}$.

We also make the following general observations about
the predicted asymmetry:
\begin{itemize}

\item As a consequence of the differing signatures
for the Pomeron and Odderon,
there is no interference between the two exchanges if they have the
same power $\alpha(t)$ since then
$\sin \left[\frac{\pi}{2}\left(\alpha_{\odd}(t)-
\alpha_{\pom}(t)\right)\right]=0$.
In fact, in a perturbative calculation at tree-level
the interference would be zero
in the high-energy limit $s\gg|t|$ since the two- and three-gluon exchanges
are purely imaginary and real respectively. This should be compared with
the analogous QED process, $\gamma Z \to \ell^+\ell^- Z$, where the
interference of the one- and two-photon exchange amplitudes
can explain~\cite{Brodsky2} the observed lepton asymmetries, energy
dependence, and nuclear target dependence of the experimental
data~\cite{Ting} for large angles. The asymmetry is in
the QED case proportional to the opening angle such that it
vanishes in the limit $s\gg|t|$.

\item In general, photon exchange will also contribute to the asymmetry
since the photon and the Odderon have the same quantum numbers.
The size of the photon exchange amplitude is 
$\frac{2}{3} \frac{e^2}{t}F_p(t)$ where $F_p$ is the 
proton form-factor and 2/3 is the charm quark electric charge.
The relative size of the photon and Odderon contributions 
will be discussed below when we give numerical estimates.

\item The overall sign of the asymmetry is not predicted by Regge theory.
(The sign of the Odderon amplitude is unknown.)  However,
the pole at $\alpha_\odd=1$ leads to the asymmetry having different
sign for $\alpha_{\odd}(t) < 1$ and  $\alpha_{\odd}(t)>1$ respectively.
Thus, if the Odderon intercept is larger than one,
which however is not supported by recent theoretical 
developments~\cite{Braun,Wosieck,Gauron2}, then the
asymmetry will change sign for some larger $t$ where $\alpha_{\odd}(t)$
goes through 1.

\end{itemize}

The ratio of the Odderon and Pomeron couplings to the proton,
$g_{pp^\prime}^{\odd}/ g_{pp^\prime}^{\pom}$,
is limited by data on
the difference of the elastic proton-proton and proton-antiproton
cross-sections at large energy $s$.
Following~\cite{kilian_nachtmann} we use the estimated limit on
the difference between the ratios of the real and imaginary part
of the proton-proton and proton-antiproton forward amplitudes,
\begin{equation}
|\Delta \rho(s)| =
\left|\frac{\Re \{ {\cal M}^{pp}(s,t=0)\} }
{\Im \{ {\cal M}^{pp}(s,t=0)\} } -
\frac{\Re \{ {\cal M}^{p\bar{p}}(s,t=0)\} }
{\Im \{ {\cal M}^{p\bar{p}}(s,t=0)\} }
\right|
\leq 0.05
\end{equation}
for $s\sim10^4$ GeV$^2$ to get a limit on the ratio of the
Odderon and Pomeron couplings to the proton.
Using the amplitude corresponding to Eq.~(\ref{eq:ampl}) for proton-proton
and proton-antiproton scattering we get for $t=0$,
\begin{eqnarray}
\Delta \rho(s) & = &
 2\frac{\Re \{ {\cal M}^{\odd}(s)\} }
       {\Im \{ {\cal M}^{\pom}(s)\} + \Im \{ {\cal M}^{\odd}(s)\}}
\simeq -2\left(\frac{g_{pp^\prime}^{\odd}}{ g_{pp^\prime}^{\pom}} \right)^2
\left(\frac{s}{s_0}\right)^{\alpha_{\odd} -\alpha_{\pom}}
\tan{\frac{\pi\alpha_\odd}{2}} ,
\end{eqnarray}
where $s_0$ is a typical hadronic scale $\sim 1$ GeV$^2$
which replaces $M_X^2$ in Eq.~(\ref{eq:ampl}). In the last step we
also make the simplifying assumption that the contribution to the denominator
from the Odderon is numerically much smaller
than from the Pomeron and therefore can be neglected.
The maximally allowed Odderon coupling at t=0 is then given by,
\begin{equation}\label{eq:rholim}
\left|g_{pp^\prime}^{\odd}\right|_{\max} =
\left|g_{pp^\prime}^{\pom}\right|
\sqrt{\frac{\Delta \rho_{\max}(s)}{2} \cot \frac{\pi\alpha_{\odd}}{2}
\left(\frac{s}{s_0}\right)^{\alpha_{\pom} -\alpha_{\odd}}} .
\end{equation}
Strictly speaking this limit applies for the soft Odderon and Pomeron and 
is therefore not directly applicable to charm photoproduction which 
is a harder process, {\it i.e.} with larger energy dependence. According to
recent data  from HERA~\cite{HERAdata} the energy dependence, parameterized as 
$s_{\gamma p}^\delta$, for photoproduction of $J/\psi$ 
mesons is $\delta=0.39\pm0.09$ for exclusive production and
$\delta=0.45\pm0.13$ for inclusive production corresponding 
to a Pomeron intercept of $\alpha_{\pom}(0)\simeq1.2$.
Even so we will use this limit to get an estimate of 
the maximal Odderon coupling to the proton.

\begin{figure}[htb]
\begin{center}
\leavevmode
\epsfbox{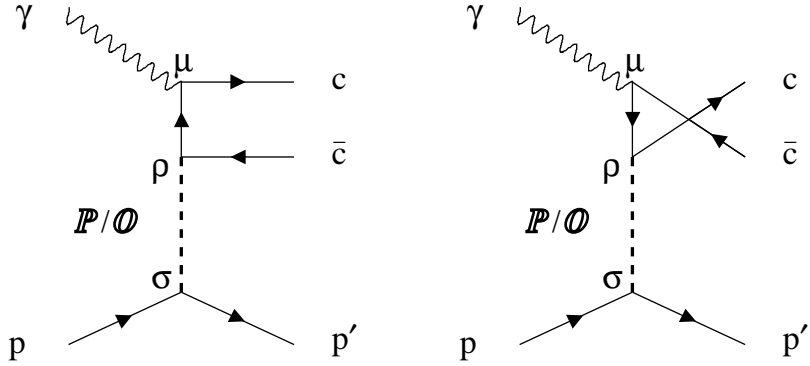}
\end{center}
\caption[*]{The amplitudes for the asymmetry
using the Donnachie-Landshoff\protect\cite{donnachie_landshoff}
 model for the Pomeron/Odderon coupling to the quark and the proton.}
\label{fig:prf}
\end{figure}

\begin{figure}[htb]
\begin{center}
\mbox{\epsfig{figure=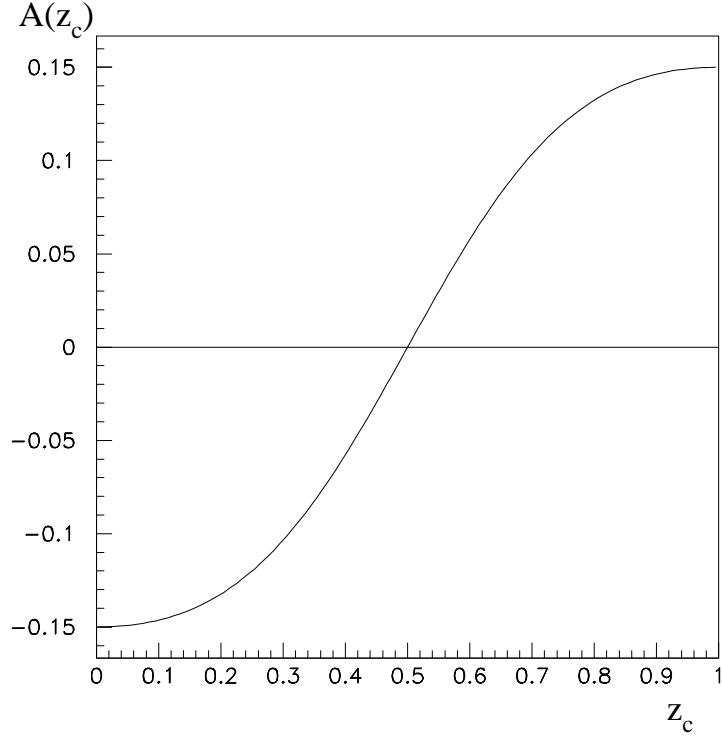,width=10cm}}
\end{center}
\caption[*]{The asymmetry in fractional energy $z_c$ of charm versus
anticharm jets predicted by our model using the
Donnachie-Landshoff Pomeron for $ \alpha_{\pom} = 1.2$,
$\alpha_{\odd} = 0.95$ and ${s_{\gamma p}}/{M_X^2}=100$.}
\label{fig:asym}
\end{figure}

The amplitudes
can be calculated using the Donnachie-Landshoff~\cite{donnachie_landshoff}
model for the Pomeron and a similar ansatz for the
Odderon~\cite{kilian_nachtmann}.
The coupling of the Pomeron/Odderon to a quark is then given by
$\kappa^{\gamma c\bar{c}}_{\pom/\odd}\gamma^\rho$,
{\it i.e.} assuming a helicity preserving local interaction.
In the same way the Pomeron/Odderon couples to the proton with
$3\kappa_{pp^\prime}^{\pom/\odd}F_1(t)\gamma^\sigma$ if we only include 
the Dirac form-factor $F_1(t)$.
The amplitudes
shown in Fig.~\ref{fig:prf} can then be obtained by replacing
$g_{pp^\prime}^{\pom/\odd}(t)g_{\pom/\odd}^{\gamma c\bar{c}}(t,M_X^2,z_c)$
in Eq.~(\ref{eq:ampl}) by,
\begin{eqnarray*}
g_{pp^\prime}^{\pom/\odd}(t) g_{\pom/\odd}^{\gamma c\bar{c}}(t,M_X^2,z_c)
 & = &
3\kappa_{pp^\prime}^{\pom/\odd}F_1(t)\bar{u}(p-\ell)\gamma^{\sigma}u(p)
\left(g^{\rho\sigma} -
\frac{\ell^\rho q^\sigma  + \ell^\sigma q^\rho }{\ell q}\right)
\kappa^{\gamma c\bar{c}}_{\pom/\odd}\epsilon^{\mu}(q)
\nonumber \\ && \times
\bar{u}(p_c)\left\{
\gamma^\mu \frac{\not\ell \,- \not p_{\bar{c}}+m_c}{(1-z)M_X^2}\gamma^\rho
-S_{\pom/\odd}
\gamma^\rho \frac{\not p_{c} \,- \not \ell+m_c}{zM_X^2}\gamma^\mu
\right\}v(p_{\bar{c}})
\end{eqnarray*}
where $\ell=\xi p$ is the Pomeron/Odderon momentum and
$g^{\rho\sigma} -\frac{\ell^\rho q^\sigma  + \ell^\sigma q^\rho }{\ell q}$
stems from the Pomeron/Odderon ``propagator". Note the signature
which is inserted for the crossed diagram to model the charge
conjugation property of the Pomeron. The Pomeron amplitude written
this way is not gauge invariant and therefore we use radiation
gauge also for the photon, {\it i.e.} the polarization sum is obtained using
$g^{\mu\nu}-\frac{q^\mu p^\nu + q^\nu p^\mu}{pq}$ (for a thorough
analysis of the gauge-dependence of the Pomeron model see~\cite{Diehl}).
The leading terms in a $t/M_X^2$ expansion of the
squared amplitudes for the Pomeron and Odderon exchange as well
as the interference are then given by,
\begin{eqnarray}
\left(\frac{g_{pp^\prime}^{\pom}g_{\pom}^{\gamma c\bar{c}}}
{\kappa_{pp^\prime}^{\pom}\kappa_\pom^{\gamma c\bar{c}}}\right)^2
& \propto &
\frac{z_c^2+z_{\bar{c}}^2}{z_cz_{\bar{c}}}\frac{(1-\xi)}{\xi^2}
\nonumber \\
\left(\frac{g_{pp^\prime}^{\odd}g_{\odd}^{\gamma c\bar{c}}}
{\kappa_{pp^\prime}^{\odd}\kappa_\odd^{\gamma c\bar{c}}}\right)^2
& \propto &
\frac{z_c^2+z_{\bar{c}}^2}{z_cz_{\bar{c}}}\frac{(1-\xi)}{\xi^2}
\nonumber \\
\frac{g_{pp^\prime}^{\pom}g_{pp^\prime}^{\odd}g_{\pom}^{\gamma c\bar{c}}
g_{\odd}^{\gamma c\bar{c}}}
{\kappa_{pp^\prime}^{\pom}\kappa_{pp^\prime}^{\odd}
\kappa_\pom^{\gamma c\bar{c}}\kappa_\odd^{\gamma c\bar{c}}}
& \propto &
\frac{z_c-z_{\bar{c}}}{z_cz_{\bar{c}}}\frac{(1-\xi)}{\xi^2} ,
\end{eqnarray}
with corrections that are of order $t/M_X^2$ and therefore can be safely
neglected. The ratio
between the interference term and the Pomeron exchange is thus given by,
\begin{equation}\label{eq:upper}
\frac{g_{pp^\prime}^{\odd}g_{\odd}^{\gamma c\bar{c}}}
     {g_{pp^\prime}^{\pom}g_{\pom}^{\gamma c\bar{c}}}
=\frac{\kappa_{pp^\prime}^{\odd}\kappa_{\odd}^{\gamma c\bar{c}}}
{\kappa_{pp^\prime}^{\pom}\kappa_\pom^{\gamma c\bar{c}}}
\frac{z_c-z_{\bar{c}}}{z_c^2+z_{\bar{c}}^2}
=\frac{\kappa_{pp^\prime}^{\odd}\kappa_{\odd}^{\gamma c\bar{c}}}
{\kappa_{pp^\prime}^{\pom}\kappa_\pom^{\gamma c\bar{c}}}
\frac{2z_c-1}{z_c^2+(1-z_c)^2}
\end{equation}
Inserting this into the asymmetry given by Eq.~(\ref{eq:pred})
and making the simplifying assumption that the
Odderon contribution can be dropped in the denominator gives
\begin{eqnarray}\label{eq:simp}
{\cal A}(t,M_X^2,z_c) & \simeq &
2
\frac{\kappa_{pp^\prime}^{\odd}\kappa_{\odd}^{\gamma c\bar{c}}}
{\kappa_{pp^\prime}^{\pom}\kappa_\pom^{\gamma c\bar{c}}}
\sin \left[\frac{\pi\left(\alpha_{\odd}-\alpha_{\pom}\right)}{2}\right]
\left(\frac{s_{\gamma p}}{M_X^2}\right)^{\alpha_{\odd}-\alpha_{\pom}}
\frac{\sin{ \frac{\pi\alpha_\pom}{2}}}{\cos{ \frac{\pi\alpha_\odd}{2}}}
\frac{2z_c-1}{z_c^2+(1-z_c)^2} \;.
\end{eqnarray}
To obtain a numerical estimate of the asymmetry, we
shall assume that $t\simeq 0$ and use
$ \alpha_{\pom}^{hard} = 1.2$  and $\alpha_{\odd} = 0.95$
\cite{Wosieck} for the Pomeron and Odderon intercepts respectively.
In addition we will also assume
${\kappa_{\odd}^{\gamma c\bar{c}} }/{\kappa_\pom^{\gamma c\bar{c}} } 
\sim \sqrt{C_F\alpha_s(m_c^2)} \simeq 0.6$, in analogy to the couplings which 
occur in the higher order corrections to Bethe-Heitler pair
production~\cite{Bethe}, and use the maximal Odderon-proton coupling,
$\kappa_{pp^\prime}^{\odd}/\kappa_{pp^\prime}^{\pom} =
g_{pp^\prime}^{\odd}/g_{pp^\prime}^{\pom}=0.1$,
which follows from Eq.~(\ref{eq:rholim}) for 
$ \alpha_{\pom}^{soft} = 1.08$, $s=10^4$ GeV$^2$,
$s_0=1$ GeV$^2$ and $\Delta \rho_{\max}(s)=0.05$.
Inserting the numerical values discussed above then gives
\begin{equation}
{\cal A} (t\simeq 0,M_X^2,z_c) \simeq 0.45 \;
\left(\frac{s_{\gamma p}}{M_X^2}\right)^{-0.25} \;
\frac{2z_c-1}{z_c^2+(1-z_c)^2}\; ,
\end{equation}
which for a typical value of $\frac{s_{\gamma p}}{M_X^2}=100$ becomes
a $\sim15$ \% asymmetry for large $z_c$ as illustrated in Fig.~\ref{fig:asym}.
We also note that the asymmetry can be integrated over $z_c$
giving
\begin{equation}
{\cal A} (t\simeq 0,M_X^2)=\int_{0.5}^{1}{\cal A} (t\simeq 0,M_X^2,z_c)
-\int_{0}^{0.5}{\cal A} (t\simeq 0,M_X^2,z_c)
\simeq 0.3\left(\frac{s_{\gamma p}}{M_X^2}\right)^{-0.25}.
\end{equation}

It should be emphasized that the magnitude of this estimate is quite
uncertain. The Odderon coupling to the proton which we are using is
a maximal coupling for the soft Odderon in relation to the soft Pomeron. 
So on the one hand the ratio may be smaller than this, 
and on the other hand the ratio may be larger 
if the hard Odderon and Pomeron have a different ratio for the
coupling to the proton. For the hard Pomeron the coupling 
is in general different at the two vertices (see e.g.~\cite{bhs})
and this could also be true for the hard Odderon.

There is also a small irreducible asymmetry from photon-Pomeron interference.
Adding the photon exchange amplitude to the Odderon amplitude modifies
the asymmetry as follows (again only taking into account the Dirac 
form-factor), 
\begin{eqnarray*}
{\cal A} (t\simeq 0,M_X^2,z_c) &=&
2 \frac{\sin{ \frac{\pi\alpha_\pom}{2}}}{{ \kappa_{pp^\prime}^{\pom}
\kappa_{\pom}^{\gamma c\bar{c}}}}  
\left(\frac{s_{\gamma p}}{M_X^2}\right)^{1-\alpha_{\pom}}
\frac{2z_c-1}{z_c^2+(1-z_c)^2} \\ &&
\left\{
-\kappa_{pp^\prime}^{\odd}
\kappa_{\odd}^{\gamma c\bar{c}} 
\left(\frac{s_{\gamma p}}{M_X^2}\right)^{\alpha_{\odd}-1 }
\frac{2\sin \left[\frac{\pi}{2}\left(\alpha_{\odd}-\alpha_{\pom}\right)
 \right]}
{\pi( \Delta_\odd+\alpha_\odd^\prime t)}
+ \frac{1}{3} \frac{2}{3} \frac{e^2}{t} \cos \frac{\pi\alpha_\pom}{2}
 \right\}
\;.
\end{eqnarray*}
where $\alpha_\odd = 1 + \Delta_\odd+\alpha_\odd^\prime t$ has been used 
to expand the pole-factor for the Odderon, 
$\cos{ \frac{\pi\alpha_\odd}{2}}
\simeq -\frac{\pi}{2}(\Delta_\odd+\alpha_\odd^\prime t)$, for small $t$.
Note that if $\Delta_\odd=0$ then the Odderon amplitude appears to have 
a $1/t$ pole just as photon exchange. However this pole must be screened
by an effective mass for the corresponding 3-gluon state.
The extra factor $1/3$ for photon exchange reflects the relative factor
of $3$ for the Pomeron/Odderon couplings to the proton~\cite{landshofforg}.
Using the soft Pomeron-proton coupling~\cite{donnachie_landshoff} 
to estimate $ \kappa_{pp^\prime}^{\pom}\kappa_{\pom}^{\gamma c\bar{c}} 
/\sin{ \frac{\pi\alpha_\pom}{2}} \simeq 3.4$ GeV$^{-2}$ gives the 
minimal asymmetry from photon-Pomeron interference (neglecting the Odderon 
contribution),
\begin{eqnarray*}
{\cal A}^{\gamma\pom} (t\simeq 0,z_c) \simeq 
-\frac{0.002}{t}\frac{2z_c-1}{z_c^2+(1-z_c)^2} \mbox{GeV}^2 \; ,
\end{eqnarray*}
where we again have used $\frac{s_{\gamma p}}{M_X^2}=100$ and $\alpha_\pom=1.2$.
Thus for very small $t$ the photon-Pomeron interference can be sizeable,
but for larger $t$ it is presumably negligible compared to Odderon-Pomeron 
interference.

In specific models, such as diquark clustering in the  proton~\cite{rueter},
the Odderon coupling to the proton in diffractive dissociation  is expected to
be larger. In such a scenario the asymmetry from  Odderon-Pomeron interference
will be correspondingly larger for proton dissociation.

In summary we have presented a sensitive test for detecting the separate
existence of the Pomeron and the Odderon exchange contributions in the
high-energy limit $s\gg|t|$ as predicted by QCD.
By observing the charge asymmetry of the quark/antiquark energy fraction ($z_c$)
in diffractive $c\bar{c}$ pair photoproduction, the interference
between the Pomeron and the Odderon exchanges can be isolated, and the
ratio to the sum of the Pomeron and the Odderon exchanges can be measured.
Using a  model with helicity conserving coupling for the Pomeron/Odderon 
to quarks, the asymmetry is predicted to be proportional to 
$(2z_c-1)/(z_c^2+(1-z_c)^2)$. The magnitude of the asymmetry is estimated
to be of order 15\%. However, this estimate includes several unknowns and
is thus quite uncertain. Such a test  could be performed
by   current experiments  at HERA and possibly COMPASS measuring the 
diffractive production of open charm in photoproduction or electroproduction.
Such measurements could provide the first experimental evidence for the
existence of the Odderon,  as well as the relative
strength of the Odderon and Pomeron couplings. Most important, the energy
dependence of the asymmetry can be used to determine  whether the Odderon
intercept is in fact greater or less than that of the Pomeron.

\section*{Acknowledgments}
C. M. thanks the Theory Group at SLAC for their kind hospitality and
Prof.~C.~Pajares of the University of Santiago de Compostela, the
Director  of the research project which partially financed this work.
We would also like to thank Markus Diehl for conversations.

\end{document}